\documentclass[a4paper,11pt]{article}

\usepackage{jheppub}

\usepackage{amsmath,amssymb,braket,longtable}
\usepackage{graphicx}
\usepackage{color}

\preprint{%
  \begin{flushright}
    UTHEP-726 \\ UTCCS-P-117 \\ KANAZAWA-18-07
  \end{flushright}
}

\title{Tensor network analysis of critical coupling \\ in two dimensional $\phi^{4}$ theory}

\author[a,b]{Daisuke Kadoh,}
\author[c]{Yoshinobu Kuramashi,}
\author[d]{Yoshifumi Nakamura,}
\author[e]{Ryo Sakai,}
\author[e]{Shinji Takeda,}
\author[c]{and Yusuke Yoshimura}

\affiliation[a]{Department of Physics, Faculty of Science, Chulalongkorn University, Bangkok 10330, Thailand}
\affiliation[b]{Research and Educational Center for Natural Sciences, Keio University, Yokohama 223-8521, Japan}
\affiliation[c]{Center for Computational Sciences, University of Tsukuba, Tsukuba 305-8577, Japan}
\affiliation[d]{RIKEN Center for Computational Science, Kobe 650-0047, Japan}
\affiliation[e]{Institute for Theoretical Physics, Kanazawa University, Kanazawa 920-1192, Japan}

\emailAdd{kadoh@keio.jp}
\emailAdd{kuramasi@het.ph.tsukuba.ac.jp}
\emailAdd{nakamura@riken.jp}
\emailAdd{sakai@hep.s.kanazawa-u.ac.jp}
\emailAdd{takeda@hep.s.kanazawa-u.ac.jp}
\emailAdd{yoshimur@ccs.tsukuba.ac.jp}

\abstract{%
  We make a detailed analysis of the spontaneous $Z_{2}$-symmetry breaking in the two dimensional real $\phi^{4}$ theory
  with the tensor renormalization group approach, which allows us to take
  the thermodynamic limit easily and determine the physical observables without statistical uncertainties.
  We determine the critical coupling in the continuum limit
  employing the tensor network formulation
  for scalar field theories proposed in our previous paper.
  We obtain $\left[ \lambda / \mu_{\mathrm{c}}^{2} \right]_{\mathrm{cont.}} = 10.913(56)$ with the quartic coupling $\lambda$ and the renormalized critical mass $\mu_{\mathrm{c}}$.
  The result is compared with previous results obtained by different approaches.
}

\begin{document}
\maketitle
\flushbottom

\section{Introduction}
\label{sec:Introduction}

The two dimensional real $\phi^{4}$ theory has been extensively studied as the simplest model
with a spontaneous symmetry breaking which plays important roles
in particle physics.
Although the continuous symmetry as in the complex $\phi^{4}$ theory cannot be broken
in two dimensions~\cite{Coleman:1973ci},
the real $\phi^{4}$ theory has a discrete $Z_{2}$-symmetry
that can be spontaneously broken~\cite{Chang:1976ek}.
The critical coupling, which separates two phases, has been studied by various approaches with and without lattice simulations~\cite{Chang:1976ek,Funke:1987wb,Harindranath:1987db,Harindranath:1988zt,Polley:1989wf,Bender:1992yd,Hauser:1994mb,Sugihara:1997xh,Lee:1998hs,Marrero:1999gf,Lee:2000ac,Hansen:2002zt,Vinnikov:2002ce,Sugihara:2004qr,Schaich:2009jk,Wozar:2011gu,Milsted:2013rxa,Rychkov:2014eea,Bosetti:2015lsa,Pelissetto:2015yha,Burkardt:2016ffk,Bronzin:2018tqz}.
Further improvements will be achieved in lattice simulations if one can reach the thermodynamic limit far beyond the system size the conventional Monte Carlo methods can access to.

The tensor renormalization group (TRG)~\cite{Levin:2006jai}
enables us to take the thermodynamic limit easily:
the computational cost of the TRG depends on the space-time volume logarithmically
while that of the Monte Carlo methods is proportional to the volume.
Furthermore, the TRG method does not contain any stochastic procedure to evaluate physical quantities.
Then, using the TRG, we can observe precisely the symmetry breaking effects with larger volumes avoiding statistical uncertainties.

Since the TRG was originally proposed for the two dimensional Ising model,
several improved algorithms~\cite{Gu:2009dr,2015PhRvL.115r0405E,yang2017loop}
and extensions to fermion systems~\cite{Gu:2010yh,Gu:2013gba}
have been developed.
Those techniques are introduced into the path-integral formulation of the field theory
and many studies have been carried out~\cite{Liu:2013nsa,Yu:2013sbi,Denbleyker:2013bea,Shimizu:2014uva,Unmuth-Yockey:2014afa,Shimizu:2014fsa,Shimizu:2017onf,Takeda:2014vwa,Kawauchi:2016xng,Meurice:2016mkb,Sakai:2017jwp,Yoshimura:2017jpk,Kuramashi:2018mmi}.
It is, however, not straightforward to apply the TRG to lattice scalar theories
because the indices of tensor are essentially the field variables so that
their dimension is infinite for scalar fields unlike the Ising case.~\footnote{
  In lattice gauge theories, one can use
  the character expansion with truncation to construct a finite dimensional tensor.
}

Recently we have proposed a new tensor network (TN) formulation for constructing a finite dimensional tensor of scalar theories~\cite{Kadoh:2018hqq}, where we directly discretize the path-integral expression of the scalar field theory using the Gaussian quadrature formula instead of the orthonormal functions proposed in the previous work~\cite{Shimizu:2012wfa}.
As the order of Gauss--Hermite quadrature is increased, we can properly reproduce the partition function of a two dimensional free scalar theory.
The method is already well established in SUSY quantum mechanics with an interacting scalar~\cite{Kadoh:2018ele},
and two dimensional theories with interacting scalars now fall within the scope of application of this method.

In this paper we determine the critical coupling of the two dimensional lattice $\phi^{4}$ theory employing the TRG method with our TN formulation.
The critical coupling is determined from the power law behavior of the
susceptibility for
the expectation value of the scalar field near the critical point.
We extrapolate the critical coupling to the continuum limit and compare it to other recent results obtained by different approaches~\cite{Schaich:2009jk,Wozar:2011gu,Bosetti:2015lsa,Bronzin:2018tqz}.

This paper is organized as follows.
We present the two dimensional continuum and lattice $\phi^{4}$ theories to fix our notation in section~\ref{sec:2Dphi4}.
A TN formulation of the expectation value of the scalar field
is given in section~\ref{sec:TN_2Dphi4}.
In section~\ref{sec:numericalresults}
we first explain the strategy and the methodology to evaluate the critical coupling from the expectation value of the field.
The results of susceptibility are then presented in section~\ref{sec:thermodynamiclimit},
and the critical couplings at several lattice spacings are listed
in table~\ref{tab:phi4_critical} of section~\ref{sec:criticalmass}.
We show the final result of the critical coupling in the continuum limit in section~\ref{sec:continuumlimit}.
Section~\ref{sec:Summary} is devoted to summary and discussion.
In appendix~\ref{sec:TRGforZ1} the coarse-graining algorithm for the expectation value of the field is explained.
The systematic error originated from the Gaussian quadrature is discussed in appendix~\ref{sec:Kdependence}.

\section{Two dimensional $\phi^{4}$ theory}
\label{sec:2Dphi4}

The Euclidean continuum action of the two dimensional real $\phi^{4}$ theory is defined as
\begin{align}
  \label{eq:phi4_lagrangian}
  S_{\mathrm{cont.}}
  =
  \int \mathrm{d}^{2} x
  \left\{
  \frac{1}{2} \left( \partial_{\rho} \phi\left( x \right) \right)^{2}
  + \frac{\mu_{0}^{2}}{2} \phi\left( x \right)^{2}
  + \frac{\lambda}{4} \phi\left( x \right)^{4}
  \right\}
\end{align}
with a real scalar field $\phi\left(x\right) \in \mathbb{R}$, the bare mass $\mu_{0}$, and the quartic coupling constant $\lambda > 0$.
This theory is super renormalizable.
The mass renormalization is required only at one-loop level
while the renormalization of the coupling is not necessary in all orders of the perturbation theory.
The suffix zero to denote the bare parameter is used only for the mass.

This model has the $Z_{2}$-symmetry ($\phi \rightarrow -\phi$) classically, but
it may be spontaneously broken at a quantum level.
The expectation value of the scalar field $\left<\phi\left(x\right)\right>$, which does not actually depend on the coordinate $x$ from the translational invariance,
is an order parameter of the symmetry breaking:
$\left<\phi\left(x\right)\right>=0$ for the symmetric phase and $\left<\phi\left(x\right)\right> \neq 0$ for the symmetry broken phase.

The dimensionful coupling $\lambda$ gives a typical scale of this model
while a dimensionless coupling normalized by $\mu^{2}$
\begin{align}
  \kappa =\frac{\lambda}{\mu^{2}}
\end{align}
characterizes its physical behavior,
where $\mu$ is the renormalized mass.
At the tree level, two phases and the sign of $\kappa_{0}=\lambda/\mu_{0}^{2}$
are in one-to-one correspondence, and the critical point is $\kappa_{0, \mathrm{c}}=0$.
The negative one corresponds to the broken phase.
However, the critical point can change with the quantum corrections.
The lattice simulations play a crucial role to determine the renormalized critical point $\kappa_{\mathrm{c}}$
beyond the perturbation theory.

Let us define the lattice theory on a square lattice
\begin{align}
  \Gamma_{L} = \Set{\left( n_{1}, n_{2} \right) | n_{1}, n_{2} = 1, 2, \ldots, L},
\end{align}
where  a positive integer $L$ is the lattice size.
The lattice spacing $a$ is assumed to be $a=1$, and we suppress it unless necessary.
We label the lattice site $(n_{1}, n_{2})$ simply as $n$.
The lattice scalar field $\phi_{n}$ is defined on the site $n$ and satisfies the periodic boundary conditions,
\begin{align}
  \phi_{n + \hat{\rho}L} = \phi_{n} && \text{for } \rho = 1, 2,
\end{align}
where  $\hat{\rho}$ is the unit vector of the $\rho$-direction.
The lattice action is given by
\begin{align}
  \label{eq:latticeaction}
  S
  = \sum_{n \in \Gamma_{L}} \left\{
  \frac{1}{2} \sum_{\rho=1}^{2} \left( \phi_{n+\hat{\rho}} - \phi_{n} \right)^{2}
  + \frac{\mu_{0}^{2}}{2} \phi_{n}^{2}
  + \frac{\lambda}{4} \phi_{n}^{4}
  \right\}.
\end{align}
Note that $\mu_{0}$ and $\lambda$ are dimensionless quantities written as $a^{2} \mu_{0}^{2}$ and $a^{2} \lambda$, respectively.

The mass renormalization can be calculated by the lattice perturbation theory.
We employ the same renormalization condition as in refs.~\cite{Chang:1976ek,Schaich:2009jk}.
The renormalized mass $\mu$ is defined as
\begin{align}
  \label{eq:renormalizedmass}
  \mu^{2}
  = \mu_{0}^{2}
  + 3 \lambda A\left( \mu^{2} \right)
\end{align}
with
\begin{align}
  \label{eq:divergent}
  A\left( \mu^{2} \right)
  = \frac{1}{L^{2}} \sum_{k_{1} = 1}^{L} \sum_{k_{2} = 1}^{L}
  \frac{1}{\mu^{2} + 4 \sin^{2}\left( \pi k_{1} / L \right) + 4 \sin^{2} \left( \pi k_{2} / L \right)},
\end{align}
which is the one-loop self energy.
We use eq.~\eqref{eq:renormalizedmass}  to determine the renormalized critical coupling in later section.

\section{Tensor network formulation for two dimensional lattice $\phi^{4}$ theory}
\label{sec:TN_2Dphi4}

The expectation value of the scalar field can be expressed as a TN form.
We present the detailed procedure following the formulation given in
our previous work~\cite{Kadoh:2018hqq}.

We begin with the partition function
\begin{align}
  \label{eq:phi4_partitionfunction}
  &Z
    = \int \mathcal{D}\phi \ e^{-S_{h}}
\end{align}
with
\begin{align}
  \label{eq:phi4_measure}
  \int \mathcal{D}\phi =  \prod_{n \in \Gamma_L} \int_{-\infty}^{\infty} \mathrm{d}\phi_{n}.
\end{align}
The action $S_{h}$ is given by
\begin{align}
  \label{eq:S_externalfield}
  S_{h}
  = S  - h \sum_{n \in \Gamma_{L}} \phi_{n},
\end{align}
where  a constant external field $h$ is introduced to investigate the phase transition
taking $h \rightarrow 0$
after $L \rightarrow \infty$.

The Boltzmann weight can be expressed as a product of local factors,
\begin{align}
  \label{eq:Boltzmannweight_aslocalfactors}
  e^{-S_{h}}
  = \prod_{n \in \Gamma_{L}} \prod_{\rho=1}^{2} f\left( \phi_{n}, \phi_{n+\hat{\rho}} \right),
\end{align}
where
\begin{align}
  \label{eq:definition_f}
  f\left( \phi_{1}, \phi_{2} \right)
  = \exp \left\{
  - \frac{1}{2}\left( \phi_{1} - \phi_{2} \right)^{2}
  - \frac{\mu_{0}^{2}}{8}\left( \phi_{1}^{2} + \phi_{2}^{2} \right)
  - \frac{\lambda}{16} \left( \phi_{1}^{4} + \phi_{2}^{4} \right)
  + \frac{h}{4}\left( \phi_{1} + \phi_{2} \right)
  \right\}.
\end{align}
The function $f\left( \phi_{1}, \phi_{2} \right)$ is a symmetric matrix with the continuous indices $\phi_{1}, \phi_{2} \in \mathbb{R}$,
and the continuity makes numerical treatments hard.

To define a finite dimensional tensor, we use the Gauss--Hermite quadrature formula,
in which the integral of a target function $g\left(x\right)$ with the weight function $e^{-x^{2}}$
is approximated by a discrete sum as
\begin{align}
  \label{eq:GHquadrature}
  \int_{-\infty}^{\infty} \mathrm{d}y e^{-y^{2}} g\left( y \right)
  \approx \sum_{\alpha=1}^{K} w_{\alpha} g\left( y_{\alpha} \right),
\end{align}
where $y_{\alpha}$ ($\alpha = 1, 2, \ldots , K$) is the $\alpha$-th root of the $K$-th Hermite polynomial $H_{K}\left(y\right)$
and
$w_{\alpha}$ is the associated weight.~\footnote{
  The $n$-th Hermite polynomial is defined by
  $H_{n}\left(x\right) = \left(-1\right)^{n} \exp \left( x^{2} \right) \left( \mathrm{d}^{n} / \mathrm{d}x^{n} \right) \exp \left( -x^{2} \right)$.
  The root $y_{\alpha}$ and weight $w_{\alpha}$ of the $n$-point Gauss--Hermite quadrature
  are given by $H_{n}\left(y_{\alpha}\right)=0$ and
  $
  w_{\alpha}= 2^{n-1} n! \sqrt{\pi} / ( n^{2} H_{n-1}\left(y_\alpha\right)^{2} )
  $.
  See ref.~\cite{abramowitz1965handbook} for the detail.
}
$K$ is the order of approximation.
If $g\left(x\right)$ is a polynomial function with the degree of $2K - 1$ or less,
the Gauss--Hermite quadrature is known to reproduce the exact result.
For general functions, the convergence of the Gauss--Hermite quadrature is not obvious,
and in this paper we numerically check the convergence for our particular case.

By replacing each integral in eq.~\eqref{eq:phi4_measure} by eq.~\eqref{eq:GHquadrature} and using eq.~\eqref{eq:Boltzmannweight_aslocalfactors},
we obtain a discretized version of the partition function
\begin{align}
  \label{eq:Z_withdiscretizedfields}
  Z\left( K \right)
  = \sum_{\left\{ \alpha \right\}} \prod_{n \in \Gamma_{L}}  w_{\alpha_{n}} \exp\left(y_{\alpha_{n}}^{2}\right)\prod_{\rho=1}^{2}
  f\left( y_{\alpha_{n}}, y_{\alpha_{n+\hat{\rho}}} \right),
\end{align}
where $\sum_{\left\{ \alpha \right\}}$ denotes $\prod_{n \in \Gamma_L}\sum_{\alpha_{n}=1}^{K}$.
This equation can be regarded as a TN representation with the bond dimension $K$. However, truncation of the bond dimension is practically required for the latter procedures because the computational cost of the TRG is more expensive than that of the Gauss--Hermite quadrature.
Then we decompose the $K\times K$ matrix $M_{\alpha\beta} = f\left(y_{\alpha}, y_{\beta}\right)$ using the singular value decomposition (SVD):~\footnote{
  The Takagi factorization is also available since $M_{\alpha\beta}$ is a square symmetric matrix.
  Then we can take $V^{\dag}=U^{\mathrm{T}}$.
}
\begin{align}
  \label{eq:SVD_f}
  f\left(y_{\alpha}, y_{\beta}\right)
  = \sum_{i=1}^{K} U_{\alpha i} \sigma_{i} V^{\dagger}_{i \beta},
\end{align}
where $\sigma_{i}$ are the singular values sorted as $\sigma_{1}\ge\sigma_{2}\ge \cdots \ge \sigma_{K}\ge0$ and
$U, V$ are unitary matrices.
The range of this summation is truncated.
We will discuss this point at the end of this section.
Finally we have
\begin{align}
  \label{eq:Z_tensornetwork}
  Z\left( K \right)
  =
  \sum_{\left\{x, t\right\}}
  \prod_{n \in \Gamma_{L}}T\left(K\right)_{x_{n} t_{n} x_{n-\hat{1}} t_{n-\hat{2}}},
\end{align}
where $\sum_{\left\{ x, t \right\}}$ means $\prod_{n \in \Gamma_L} \sum_{x_{n}=1}^{K}\sum_{t_{n}=1}^{K}$
and
\begin{align}
  \label{eq:definition_tensor}
  T\left(K\right)_{i j k l}
  = \sqrt{\sigma_{i} \sigma_{j} \sigma_{k} \sigma_{l}} \sum_{\alpha =1}^{K} w_{\alpha}
  e^{y_{\alpha}^{2}}
  U_{\alpha i} U_{\alpha j} V^{\dagger}_{k \alpha}  V^{\dagger}_{l \alpha}.
\end{align}

Now let us turn to the expectation value of $\phi$,
\begin{align}
  \label{eq:expectationvalue_phi}
  \left< \phi_{n} \right>
  = \frac{Z_{1}}{Z},
\end{align}
where
\begin{align}
  \label{eq:definition_Z1}
  Z_{1}
  =   \int \mathcal{D}\phi \  \phi_{{n}} \, e^{-S_{h}}.
\end{align}
Since $Z$ is already expressed as a uniform TN as shown in eq.~\eqref{eq:Z_tensornetwork},
the remaining task is to represent the numerator $Z_{1}$ as a TN.
The insertion of the operator $\phi_{n}$ in eq.~\eqref{eq:definition_Z1} affects the form of the tensor as
\begin{align}
  \label{eq:definition_tildeT}
  \tilde{T}\left(K\right)_{i j k l}
  = \sqrt{\sigma_{i} \sigma_{j} \sigma_{k} \sigma_{l}} \sum_{\alpha =1}^{K} y_{\alpha} w_{\alpha}
  e^{y_{\alpha}^{2}}
  U_{\alpha i} U_{\alpha j} V^{\dagger}_{k \alpha} V^{\dagger}_{l \alpha},
\end{align}
where an extra $y_{\alpha}$ is inserted compared to eq.~\eqref{eq:definition_tensor}.
We call the modified tensor $\tilde{T}$ impurity tensor.
Repeating the same procedures as for the partition function,
we obtain the following TN form:
\begin{align}
  \label{eq:Z1_tensornetwork}
  Z_{1}\left( K \right)
  = \sum_{\left\{ x, t \right\}} \tilde{T}\left(K\right)_{x_{n} t_{n} x_{n-\hat{1}} t_{n-\hat{2}}}
  \prod_{m \in \Gamma_{L}, m \neq n} T\left(K\right)_{x_{m} t_{m} x_{m-\hat{1}} t_{m-\hat{2}}},
\end{align}
in which one of $T$s in eq.~\eqref{eq:definition_tensor} is replaced by the impurity tensor $\tilde{T}$.

The contraction of tensor indices in eq.~\eqref{eq:Z_tensornetwork} is exactly taken from $1$ to $K$.
In actual computations, we initially truncate the bond dimension from $K$ to $D$ on account of computational complexity;
namely we redefine $\sum_{\left\{ x, t \right\}}$ as $\prod_{n \in \Gamma_{L}} \sum_{x_{n}=1}^{D} \sum_{t_{n}=1}^{D}$,
and then the partition function in eq.~\eqref{eq:Z_tensornetwork} is initially approximated.
In coarse-graining steps the number of tensors in the network is reduced with truncating the bond dimensions.
In this study we keep the size of tensors as $D$.~\footnote{
  See appendix~\ref{sec:TRGforZ1} for more details.
}
Now we have two parameters to control the accuracy of the approximations, the degree of the Gauss--Hermite quadrature $K$ and the size of tensors $D$,
and we will check the stability of numerical results with respect to $D$ and $K$ in appendix~\ref{sec:Kdependence}.

\section{Numerical results}
\label{sec:numericalresults}

In this section the numerical results are shown step by step.
First, the technical details related to the Gauss--Hermite quadrature and the TRG is summarized in section~\ref{sec:expectationValueOfPhi}.
In section~\ref{sec:thermodynamiclimit} we extract the susceptibility of the expectation value of $\phi$ with effectively taking the thermodynamic limit.
Using the susceptibility we determine the critical parameters in section~\ref{sec:criticalmass},
and in section~\ref{sec:continuumlimit} we take the limit $\lambda \to 0$ to obtain the critical coupling in the continuum limit.

\subsection{Methods}
\label{sec:expectationValueOfPhi}

The expectation value $\left<\phi\right>$ is obtained by the ratio
of $Z\left(K\right)$ and $Z_{1}\left(K\right)$, which are individually evaluated by the TRG method with $D \in \left[ 16, 64 \right]$. One can use the standard technique for $Z\left( K \right)$
and needs a little ingenuity for  $Z_{1}\left( K \right)$.
The coarse-graining procedure of the tensor network with an impurity tensor is shown
in refs.~\cite{2008PhRvB..78t5116G,Shimizu:2012wfa,Unmuth-Yockey:2014afa,Nakamoto2016}, and
the detailed technique is also described in appendix~\ref{sec:TRGforZ1}.

\begin{figure}[htbp]
  \centering
  \includegraphics[width=0.8\hsize]{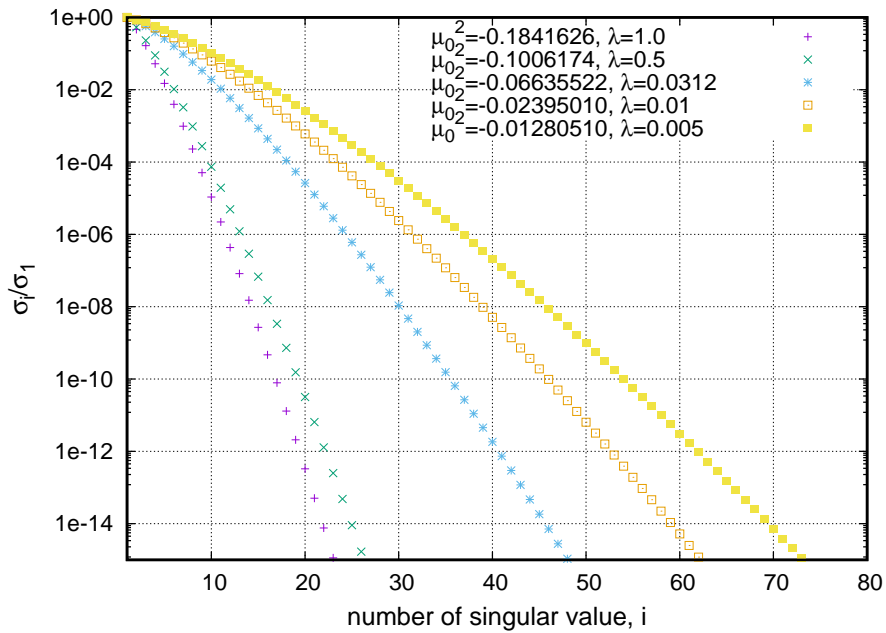}
  \caption{Singular values of the SVD in eq.~\eqref{eq:SVD_f}, which is used to define the initial tensor in eq.~\eqref{eq:definition_tensor}.}
  \label{fig:singularvalues}
\end{figure}

We remark that the systematic errors can be estimated by investigating the $K$- and $D$-dependences of the numerical results.
There are three sources of systematic errors in the present case:
(a) discretization of the scalar fields with the $K$-point Gauss--Hermite quadrature,
(b) truncation of the bond dimension of the initial tensor discussed in the last paragraph in section~\ref{sec:TN_2Dphi4},
and (c) truncation in coarse-graining steps.
The effect of (a) is appeared as $K$-dependence of the results while the errors from (b) and (c) are observed
through the $D$-dependence of the results.
In figure~\ref{fig:singularvalues} we plot the singular values of the SVD in eq.~\eqref{eq:SVD_f}.  Clear hierarchy of the singular values assures that the modification of contraction range that is explained at the end of the last section effectively work.
We fix $K = 256$ that is large enough
so that the systematic errors associated with the choice of $K$
are to be smaller than those of $D$
as discussed in appendix~\ref{sec:Kdependence}.
In section~\ref{sec:criticalmass}
we investigate the $D$-dependence of the critical coupling $\lambda / \mu_{\mathrm{c}}^{2}$
and  estimate the systematic errors by measuring small fluctuations which occur when $D$ changes.

\subsection{Susceptibility}
\label{sec:thermodynamiclimit}

The critical point is given by the peak position of the  susceptibility
for $\left<\phi\right>$  defined by
\begin{align}
  \label{eq:chi}
  \chi =
  \lim_{h\to0} \lim_{L\to \infty}
  \frac{\left< \phi \right>_{h, L}-\left< \phi \right>_{0, L}}{h},
\end{align}
where $\left< \phi \right>_{h, L}$ denotes the expectation value
measured for given constant external field $h$ and lattice size $L$.~\footnote{
  One can evaluate the susceptibility also by differentiating the free energy twice with respect to $h$ numerically,
  but it may suffer from a loss of significant digits.
  Since $\left<\phi\right>$ is obtained as a direct output of the TRG,
  we use eq.~\eqref{eq:chi} to avoid such a problem.
}
Equation~\eqref{eq:chi} is reduced to $\chi=\lim_{h \to 0} \lim_{L \to \infty} \left< \phi \right>_{h, L} / h$
since $\left< \phi \right>_{0, L} =0$.  It is rather complicated procedure to take the double limits with respect to $h$ and $L$ numerically. We first find a constant behavior of $\left< \phi \right>_{h, L} / h$
as $L$ increases, which effectively represents $\left< \phi \right>_{h, \infty} / h$.
Then, the reduction of $h$ allows us to extract a converged value corresponding to the susceptibility $\chi$.

\begin{figure}[htbp]
  \centering
  \includegraphics[width=0.8\hsize]{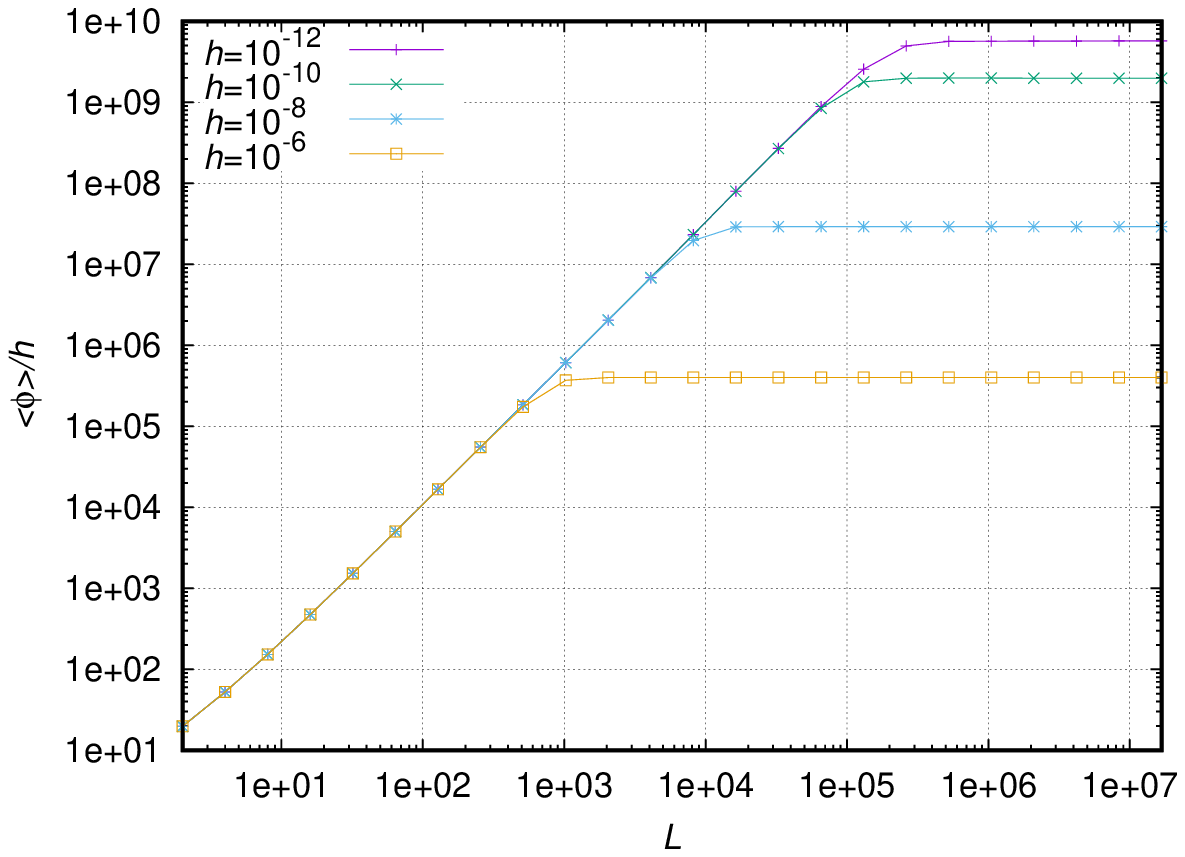}
  \caption{Thermodynamic limit of $\left< \phi \right>_{h, L}/h$ with $\mu_{0}^{2} = \mu_{0, \mathrm{rep.}}^{2}, \ \lambda=0.05,\ D=32$, and $K=256$.
  }
  \label{fig:thermodynamiclimit}
\end{figure}

In figure~\ref{fig:thermodynamiclimit} we
show $\left< \phi \right>_{h, L}/h$ as a function of $L$ for several values of $h$.
One can see that it becomes a constant for larger volumes with $L \ge 10^{6}$,
which effectively corresponds to the thermodynamic limit within the systematic errors.~\footnote{
  For sufficiently large $L$, $\left< \phi \right>_{h, L}/h$ is not exactly
  a constant but behaves as a constant within small fluctuations
  which basically come from the effects of finite $K$ and $D$.
}
$\left< \phi \right>_{h, \infty}/h$ is thus obtained for several small values of the external field $h$.
As a representative case we choose  $D=32$, $\lambda = 0.05$ and, $\mu_{0}^{2} = \mu_{0, \mathrm{rep.}}^{2}$ with
\begin{align}
  \label{eq:mu2_0rep}
  \mu_{0, \mathrm{rep.}}^{2} = -0.1006174,
\end{align}
for which the system is in the symmetric phase though it is very close to the critical point: $1 - \mu^{2}_{0, \mathrm{rep.}}/\mu^{2}_{0, \mathrm{c}} \approx 10^{-5}$.

\begin{figure}[htbp]
  \centering
  \includegraphics[width=0.8\hsize]{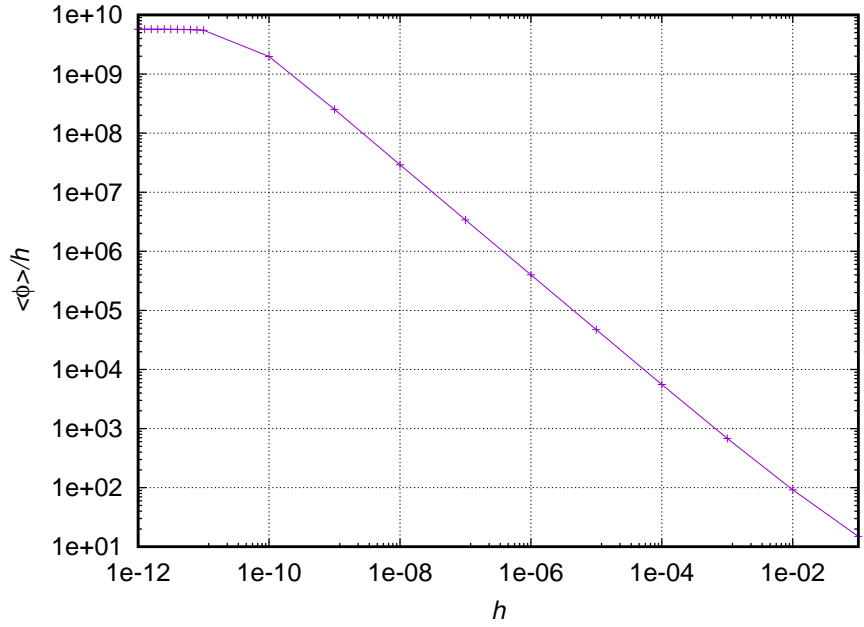}
  \caption{$\left< \phi \right>_{h, \infty}/h$ as a function of $h$ with the same parameters as in figure~\ref{fig:thermodynamiclimit}.
  }
  \label{fig:zerohlimit}
\end{figure}

\begin{figure}[htbp]
  \centering
  \includegraphics[width=0.8\hsize]{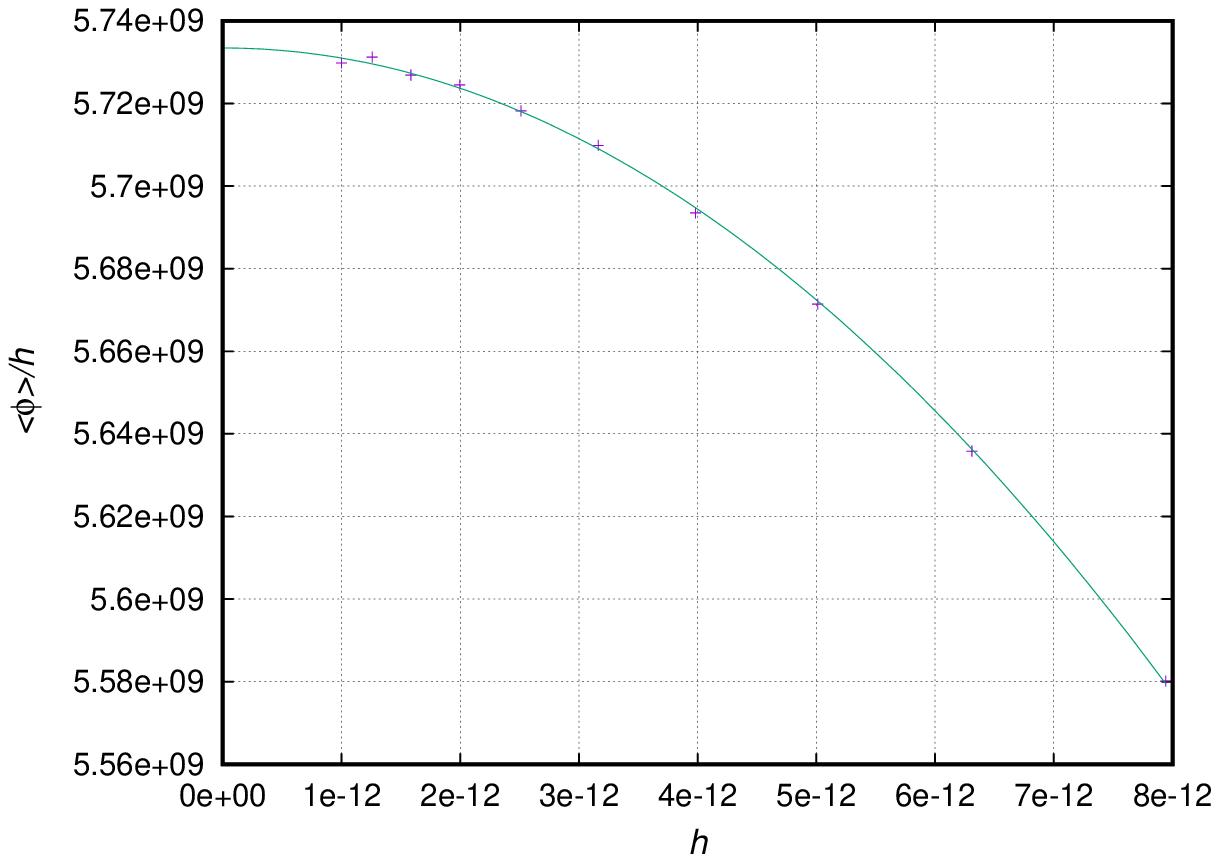}
  \caption{Closer look on $\left< \phi \right>_{h, \infty}/h$ in figure~\ref{fig:zerohlimit}.
  }
  \label{fig:zerohlimit_close}
\end{figure}

Figure~\ref{fig:zerohlimit}
shows the $h$-dependence of $\left< \phi \right>_{h, \infty}/h$ with
the same parameter set as in figure~\ref{fig:thermodynamiclimit}.
It seems to converge to a constant for sufficiently small $h \le 10^{-11}$.
We can extract the limit of $\left< \phi \right>_{h, \infty}/h$ to the vanishing $h$, which corresponds to the susceptibility $\chi$ defined in eq.~\eqref{eq:chi}
with  $\left< \phi \right>_{0, L} =0$.
Figure~\ref{fig:zerohlimit_close} shows the zoomed version of figure~\ref{fig:zerohlimit},
and there the values of $\left< \phi \right>_{h, \infty}/h$ lie on a quadratic function.
Since $\left< \phi \right>_{h, \infty}$ is an odd function of $h$, it behaves as $\left< \phi \right>_{h, \infty} = c_{1}h + c_{3}h^{3} + \cdots$ with coefficients $c_{i}$ ($i = 1, 3, \ldots$),
and one can find that the ratio does $\left< \phi \right>_{h, \infty}/h = c_{1} + c_{3}h^{2} + \cdots$.
This is the reason why the data in figure~\ref{fig:zerohlimit_close} show the quadratic behavior.
Then we simply fit them using a quadratic function to determine $\chi$,
and its error is defined as the difference between the obtained susceptibility and $\left< \phi \right>_{h, \infty}/h$ at $h=10^{-11}$.~\footnote{
  This error will propagate to the next analysis, and in the end it will become that of
  $\lambda / \mu_{\mathrm{c}}^{2}$ for given $\lambda$.
  The propagated error, however,  does not affect the final result because it is very small compared to a fluctuation originated from  $D$ as shown in figure~\ref{fig:criticalcoupling_D16-64}.
}

Similar behaviors are observed for other parameter sets employed in this work (see table~\ref{tab:phi4_critical}).  We obtain the susceptibility for all the parameter sets taking the same procedure as here.

\subsection{Critical mass for each $\lambda$}
\label{sec:criticalmass}

The scaling behavior of the susceptibility $\chi$ as a function of $\mu_{0}$ near the critical point in the symmetric phase is given by
\begin{align}
  \label{eq:scalinglaw}
  \chi = A \left| \mu_{0, \mathrm{c}}^{2} - \mu_{0}^{2} \right|^{-\gamma}
\end{align}
with the critical bare mass $\mu_{0, \mathrm{c}}$, the critical exponent $\gamma$,
and a constant $A$.
This formula can be used to determine $\mu_{0, \mathrm{c}}$ by fitting the numerical result of $\chi$.
It may be more beneficial to express the above formula
in the following way:
\begin{align}
  \label{eq:deformedfittingform}
  \chi^{-1/\gamma_{\mathrm{Ising}}}
  =A^{\prime} \left| \mu_{0, \mathrm{c}}^{2} - \mu_{0}^{2} \right|^{\gamma/\gamma_{\mathrm{Ising}}}.
\end{align}
Since the $\phi^{4}$ theory is expected to belong to the universality class of the two dimensional Ising model,
whose critical exponent is exactly known to be $\gamma_{\mathrm{Ising}} = 7/4 = 1.75$~\cite{Onsager:1943jn},
the modified expression could make it easier to check the consistency between $\gamma$ and $\gamma_{\mathrm{Ising}}$;
one can clearly see that $\gamma/\gamma_{\mathrm{Ising}} = 1$ in the right hand side of eq.~\eqref{eq:deformedfittingform}
if $\chi^{-1/\gamma_{\mathrm{Ising}}}$ behave as a straight line with respect to $\mu_{0}^{2}$ around the critical point.

\begin{figure}[htbp]
  \centering
  \includegraphics[width=0.8\hsize]{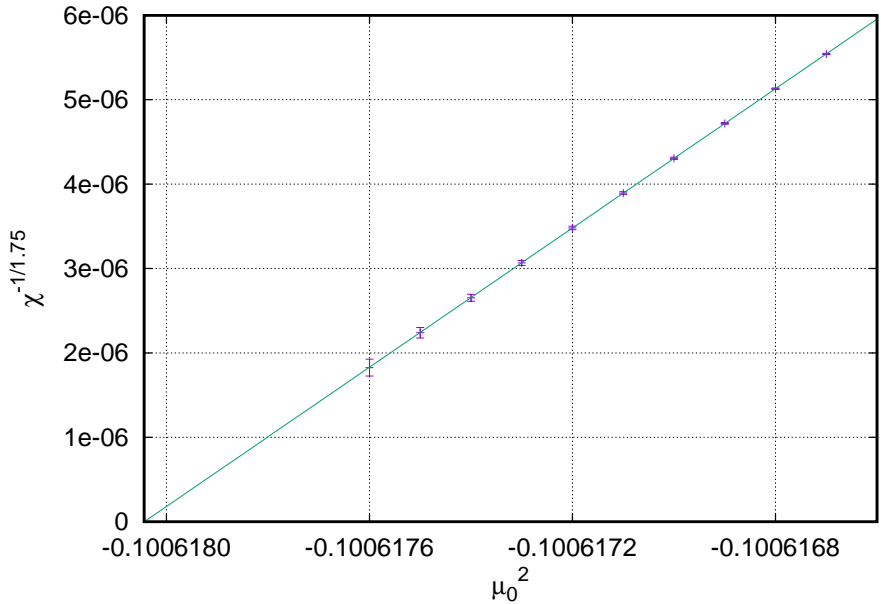}
  \caption{$\chi^{-1/1.75}$ as a function of $\mu_{0}^{2}$ at $\lambda=0.05$ and $D=32$.
  }
  \label{fig:chi_linearfit}
\end{figure}

Figure~\ref{fig:chi_linearfit} shows the susceptibility
as a function of $\mu_{0}^{2}$ at $\lambda=0.05$ and $D=32$.
It is clear that the results are on a straight line which implies $\gamma=\gamma_{\mathrm{Ising}}$.
We determine the values of  $\mu^{2}_{0, \mathrm{c}}$
by fitting the susceptibility using the expression of eq.~\eqref{eq:deformedfittingform} with $\mu^{2}_{0, \mathrm{c}}$ and $A^{\prime}$ as free parameters and fixed $\gamma = \gamma_{\mathrm{Ising}}$.

The bare critical mass $\mu_{0, \mathrm{c}}^{2}$ for another value of $\lambda$ is obtained by repeating the procedures above.~\footnote{
  In this context the critical mass is regarded as a function of $\lambda$.
  To emphasize this we do not hide the argument $\lambda$ in eq.~\eqref{eq:continuum}.
  Note that the critical mass also depends on $D$ as shown in table~\ref{tab:phi4_critical},
  but we extract the $D$-independent ones to take the continuum limit as discussed in the last of this section.
}
Then the renormalized critical mass $\mu^{2}_{\mathrm{c}}$ is determined
by solving the self-consistent equation~\eqref{eq:renormalizedmass} using $\mu^{2}_{0, \mathrm{c}}$ and $\lambda$ as inputs.

Table~\ref{tab:phi4_critical} summarizes the results of $\mu^{2}_{0, \mathrm{c}}$ obtained for $0.005\le\lambda\le0.1$ and $16\le D\le64$.
We also present the dimensionless critical coupling $\lambda/\mu_{\mathrm{c}}^{2}$.
Since $\chi^{2}/\text{d.o.f.} \leq 1$ for all $\lambda$ and $D$,
the legitimity of the linear fitting shown in figure~\ref{fig:chi_linearfit} are confirmed,
and we can consider that fixing $\gamma = \gamma_{\mathrm{Ising}}$ is a reasonable assumption.
The critical bare mass $\mu_{0, \mathrm{c}}^{2}$ monotonically decreases as $\lambda$ approaches to zero
while it does not show any smooth dependence on $D$ at each $\lambda$.

\begin{figure}[htbp]
  \centering
  \includegraphics[width=0.8\hsize,bb=0 0 504 360]{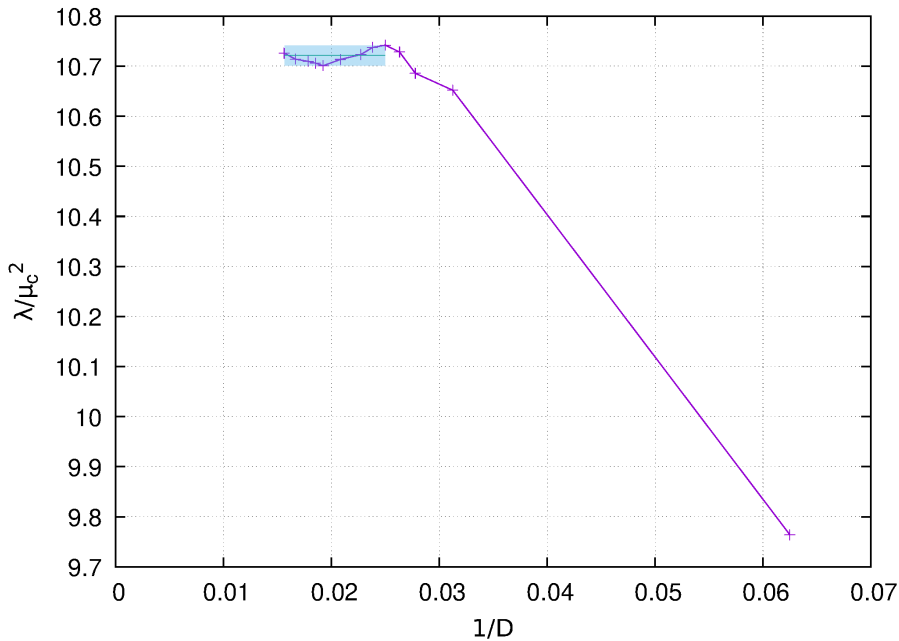}
  \caption{
    $D$-dependence of $\lambda/\mu_{\mathrm{c}}^{2}$
    for $\lambda = 0.05$.
    The error estimated from the fluctuation is also shown as the colored band.
  }
  \label{fig:criticalcoupling_D16-64}
\end{figure}

In figure~\ref{fig:criticalcoupling_D16-64} we plot the $D$-dependence of $\lambda/\mu_{\mathrm{c}}^{2}$ for $\lambda=0.05$.
In the large $D$ region $\lambda/\mu_{\mathrm{c}}^{2}$ converges to a value while oscillating.
This kind of behavior is also observed in the TRG computations of other models such as the Ising and the Potts models~\cite{2014ChPhL..31g0503W}.
As also shown in figure~\ref{fig:criticalcoupling_D16-64}
we determine the central value of  $\lambda/\mu_{\mathrm{c}}^{2}$
as an average between the maximum and minimum values in the oscillating region.
The error is estimated using the half width of the oscillation.
Although the critical coupling for each $D$ has a rather small systematic error as given in table~\ref{tab:phi4_critical},
its magnitude is negligible compared to the fluctuation of the critical coupling with the variation of $D$.
Thus,
we present the results of the renormalized critical coupling $\lambda/\mu_{\mathrm{c}}^{2}$ only with the systematic errors associated with $D$.

\setlength{\LTcapwidth}{\hsize}
{\small
  \begin{longtable}{ccccc}
    \centering
    $\lambda$ & $D$ & $\mu_{0, \mathrm{c}}^{2}$ & $\lambda/\mu_{\mathrm{c}}^{2}$ & $\chi^{2}/\text{d.o.f.}$  \\\hline\hline
    $0.1$ & $32$ & $-0.184163558(16)$ & $10.5143879(52)$ & $0.0019$ \\
    $0.1$ & $36$ & $-0.184317627(22)$ & $10.5628049(70)$ & $0.021$ \\
    $0.1$ & $40$ & $-0.184352147(32)$ & $10.573692(10)$ & $0.0015$ \\
    $0.1$ & $44$ & $-0.184356707(34)$ & $10.575132(10)$ & $0.0049$ \\
    $0.1$ & $46$ & $-0.184357220(35)$ & $10.575293(11)$ & $0.0031$ \\
    $0.1$ & $48$ & $-0.184323247(40)$ & $10.564577(12)$ & $0.0017$ \\
    $0.1$ & $52$ & $-0.184305955(40)$ & $10.559127(12)$ & $0.044$ \\
    $0.1$ & $56$ & $-0.184344222(33)$ & $10.571191(10)$ & $0.037$ \\
    \\
    $0.05$ & $32$ & $-0.1006180444(70)$ & $10.6517781(44)$ & $0.0072$ \\
    $0.05$ & $36$ & $-0.100671024(11)$ & $10.6856832(74)$ & $0.087$ \\
    $0.05$ & $38$ & $-0.100737125(13)$ & $10.7281795(89)$ & $0.08$ \\
    $0.05$ & $40$ & $-0.100758319(14)$ & $10.7418512(95)$ & $0.0051$ \\
    $0.05$ & $42$ & $-0.1007510328(81)$ & $10.7371484(52)$ & $0.87$ \\
    $0.05$ & $44$ & $-0.100729217(17)$ & $10.723084(11)$ & $0.17$ \\
    $0.05$ & $48$ & $-0.100714173(18)$ & $10.713399(11)$ & $0.015$ \\
    $0.05$ & $52$ & $-0.100694733(14)$ & $10.7009008(96)$ & $0.35$ \\
    $0.05$ & $54$ & $-0.100702523(54)$ & $10.705907(35)$ & $0.018$ \\
    $0.05$ & $56$ & $-0.100708214(17)$ & $10.709566(11)$ & $0.02$ \\
    $0.05$ & $60$ & $-0.100714918(53)$ & $10.713879(34)$ & $0.8$ \\
    $0.05$ & $64$ & $-0.100733523(19)$ & $10.725858(12)$ & $0.0078$ \\
    \\
    $0.0312$ & $32$ & $-0.066355403(12)$ & $10.698476(12)$ & $0.25$ \\
    $0.0312$ & $36$ & $-0.066424884(14)$ & $10.770365(14)$ & $0.55$ \\
    $0.0312$ & $40$ & $-0.066451867(14)$ & $10.798450(15)$ & $0.57$ \\
    $0.0312$ & $44$ & $-0.066469304(10)$ & $10.816649(10)$ & $0.11$ \\
    $0.0312$ & $48$ & $-0.066440457(12)$ & $10.786563(13)$ & $0.082$ \\
    $0.0312$ & $52$ & $-0.0664157556(89)$ & $10.7608852(92)$ & $0.16$ \\
    $0.0312$ & $56$ & $-0.066434790(23)$ & $10.780665(24)$ & $0.76$ \\
    \\
    $0.01$ & $32$ & $-0.0239502718(48)$ & $10.575644(15)$ & $0.33$ \\
    $0.01$ & $36$ & $-0.0240350323(49)$ & $10.848845(16)$ & $0.49$ \\
    $0.01$ & $40$ & $-0.0240490620(57)$ & $10.894934(18)$ & $0.26$ \\
    $0.01$ & $44$ & $-0.0240668834(51)$ & $10.953841(16)$ & $0.84$ \\
    $0.01$ & $48$ & $-0.0240397741(63)$ & $10.864395(20)$ & $0.0051$ \\
    $0.01$ & $52$ & $-0.024025540(25)$ & $10.817802(84)$ & $0.0081$ \\
    $0.01$ & $56$ & $-0.0240288600(74)$ & $10.828648(24)$ & $0.42$ \\
    \\
    $0.005$ & $36$ & $-0.0128280765(46)$ & $10.735394(29)$ & $0.4$ \\
    $0.005$ & $40$ & $-0.0128621903(37)$ & $10.958938(25)$ & $0.15$ \\
    $0.005$ & $42$ & $-0.0128734615(84)$ & $11.034101(56)$ & $0.012$ \\
    $0.005$ & $44$ & $-0.0128536408(33)$ & $10.902359(21)$ & $0.49$ \\
    $0.005$ & $48$ & $-0.0128508376(42)$ & $10.883889(27)$ & $0.37$ \\
    $0.005$ & $52$ & $-0.012841211(17)$ & $10.82077(11)$ & $0.013$ \\
    $0.005$ & $56$ & $-0.0128303006(66)$ & $10.749790(43)$ & $0.28$ \\
    $0.005$ & $60$ & $-0.0128349426(48)$ & $10.779915(31)$ & $0.0065$ \\\\
    \caption{
    Bare critical mass $\mu_{0, \mathrm{c}}^{2}$, renormalized critical coupling $\lambda/\mu_{\mathrm{c}}^{2}$, and $\chi^{2}/\text{d.o.f.}$ of the linear fitting described in the main body of the text for given $\lambda$ and $D$.
    Errors are originated from that of the susceptibility as explained in section~\ref{sec:thermodynamiclimit}.  }
    \label{tab:phi4_critical}
  \end{longtable}
}

\subsection{Continuum limit of the critical coupling}
\label{sec:continuumlimit}

We finally obtain the value of the critical coupling in the continuum limit as
\begin{align}
  \label{eq:continuum}
  \left[ \frac{\lambda}{\mu_{\mathrm{c}}^{2}} \right]_{\mathrm{cont.}}
  = \lim_{\lambda\to0}
  \frac{\lambda}{\mu_{\mathrm{c}}^{2}\left(\lambda\right)}.
\end{align}
Note that the continuum limit is understood as $a^{2} \lambda \rightarrow 0$.

\begin{figure}[htbp]
  \centering
  \includegraphics[width=0.8\hsize]{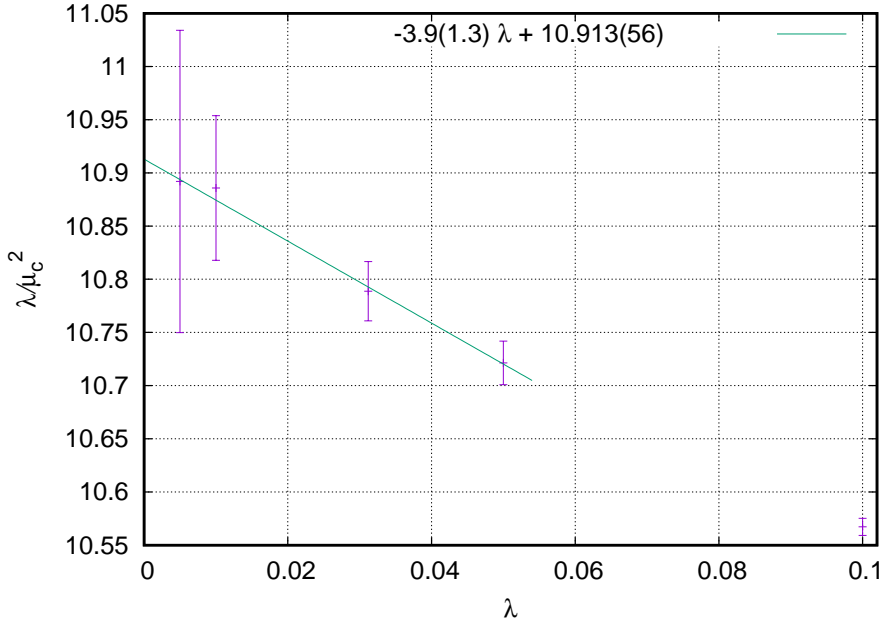}
  \caption{
    Extrapolation of $\lambda/\mu_{\mathrm{c}}^{2}$ to the continuum limit $\lambda\to0$.
    The line represents the result of the linear extrapolation with
    $\chi^{2}/\text{d.o.f.} \approx 0.026$.
  }
  \label{fig:contnuumlimit_lambda0.005-0.100_K256}
\end{figure}

Figure~\ref{fig:contnuumlimit_lambda0.005-0.100_K256} shows the $\lambda$-dependence of the critical coupling with the systematic errors.
Let us take the continuum limit ($\lambda \rightarrow 0$)
of the critical coupling by a linear extrapolation.
The result of the linear fit for $\lambda\le0.05$ shown in the figure gives a reasonable chi-squared value: $\chi^{2}/\text{d.o.f.} \approx 0.026$.
The critical coupling in the continuum limit is found to be
\begin{align}
  \label{eq:continuum_value}
  \left[ \frac{\lambda}{\mu_{\mathrm{c}}^{2}} \right]_{\mathrm{cont.}} = 10.913(56)
\end{align}
with the systematic error from the $D$-dependence in the TRG.

\begin{figure}[htbp]
  \centering
  \includegraphics[width=0.8\hsize]{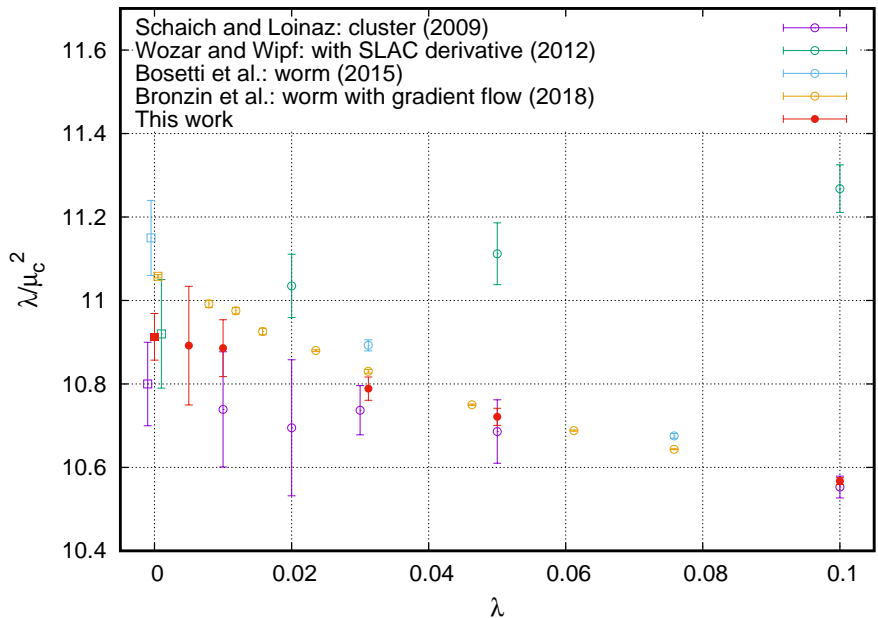}
  \caption{Comparison of the continuum extrapolations of the critical coupling $\lambda/\mu_{\mathrm{c}}^{2}$
    given in recent Monte Carlo studies (Schaich and Loinaz~\cite{Schaich:2009jk}, Wozar and Wipf~\cite{Wozar:2011gu},  Bosetti et al.~\cite{Bosetti:2015lsa},
    and Bronzin et al.~\cite{Bronzin:2018tqz}) and this work.
    At $\lambda = 0$, data points are horizontally shifted to ensure the visibility.
    Note that the results by Wozar and Wipf cannot be compared at non-zero $\lambda$
    since they used the SLAC derivative for scalar bosons,
    but, in the continuum limit ($\lambda = 0$), their results are consistent with naively discretized ones within errors.
  }
  \label{fig:continuumlimit}
\end{figure}

Figure~\ref{fig:continuumlimit} shows a comparison
among our result and the recent Monte Carlo results in refs.~\cite{Schaich:2009jk,Wozar:2011gu,Bosetti:2015lsa,Bronzin:2018tqz} for small $\lambda$.
We have reached the smallest lattice spacing: $\lambda=0.005$.
The error bar, however, is relatively large compared to the latest Monte Carlo result around $\lambda \sim 0.01$.
Note that around $\lambda = 0.0312$ one can directly compare four data points
given by different methods.
As a result, the values by Schaich and Loinaz~\cite{Schaich:2009jk},
Bronzin et al.~\cite{Bronzin:2018tqz}
and us seems roughly consistent with each other while
that by Bosetti et al.~\cite{Bosetti:2015lsa} is sizably deviated from the three results.

\section{Summary and outlook}
\label{sec:Summary}

We have studied the two dimensional lattice $\phi^{4}$ theory using the TRG with our new formulation of making a finite dimensional tensor for scalar field theories given in ref.~\cite{Kadoh:2018hqq}.
The TRG method, whose computational cost depends on the space-time volume logarithmically, allows us to access to large volume lattices so that the thermodynamic limit
and the power law behavior of the susceptibility can be precisely studied.
This work is the first study to determine the critical coupling in the continuum limit employing the TRG method.

The dimensionless critical coupling was obtained as
\begin{align}
  \label{eq:continuum_value2}
  \left[ \frac{\lambda}{\mu_{\mathrm{c}}^{2}} \right]_{\mathrm{cont.}} = 10.913(56)
\end{align}
with sufficiently small error albeit the simplest form of the TRG is employed.
The error is the systematic one coming from the finite $D$ effects.
Our result shows a reasonable consistency with the recent results obtained by different approaches.

The simplest TRG algorithm suffers from the growth of the systematic errors around  the criticality.
Alternative coarse-graining procedures such as
the tensor network renormalization (TNR)~\cite{2015PhRvL.115r0405E} and loop-TNR~\cite{yang2017loop}
might be useful to obtain more precise results.
These methods effectively work around the critical point
and, in principle, are applicable to any two dimensional model irrespective of the field contents.
We could expect further improvements for the accuracy of the critical coupling.

\appendix

\section{Coarse-graining of tensor network including impurity tensors}
\label{sec:TRGforZ1}

We describe the coarse-graining algorithm for a tensor network with an impurity tensor
such as $Z_{1}\left( K \right)$ in eq.~\eqref{eq:Z1_tensornetwork},
which is given with a fixed integer $D$ for truncating the SVD.

\begin{figure}[htbp]
  \centering
  \includegraphics[width=0.8\hsize]{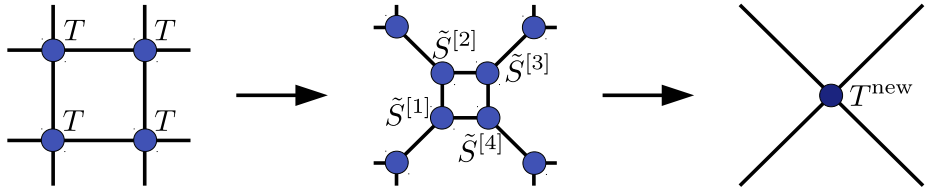}
  \caption{
    Coarse-graining step for uniform tensor network.
    Circles represent tensors,
    and closed indices are contracted.
    A local block on the network is taken into account here.
    In the first step, every rank-four tensor is decomposed into two rank-three tensors.
    By contracting the four rank-three tensors, a new rank-four tensor is obtained in the second step.
    Here, decomposed tensors with tilde are defined by $\tilde{S}^{[1(3)]} = \sqrt{\sigma^{[13]}} S^{[1(3)]}$ and $\tilde{S}^{[2(4)]} = \sqrt{\sigma^{[24]}} S^{[2(4)]}$.
  }
  \label{fig:TRG}
\end{figure}

Before discussing the nonuniform case, we first explain the coarse-graining of a uniform network
such as $Z\left( K \right)$ in eq.~\eqref{eq:Z_tensornetwork}
without impurity tensors.
The graphical representation of a coarse-graining step is given in figure~\ref{fig:TRG}.
Firstly, by using the SVD, the tensor $T$ with the bond dimension $N$ is decomposed in two ways:
\begin{align}
  \label{even_T}
  T_{ijkl} = M^{[13]}_{(ij)(kl)} = \sum_{m=1}^{N^{2}} S^{[1]}_{(ij)m} \sigma^{[13]}_{m} S^{[3]}_{m(kl)},
  \\
  \label{odd_T}
  T_{ijkl} = M^{[24]}_{(li)(jk)} = \sum_{m=1}^{N^{2}} S^{[2]}_{(li)m} \sigma^{[24]}_{m} S^{[4]}_{m(jk)},
\end{align}
where $M^{[13]}$ and $M^{[24]}$ ($N^{2}\times N^{2}$ matrices) are obtained by arranging the indices of $T$, $\sigma^{[13]}$ and $\sigma^{[24]}$ are the singular values in the descending order,  and $S^{[i]}$ ($i=1, 2, 3, 4$) denotes the singular matrix.
We actually apply eq.~\eqref{even_T} to tensors on even sites and eq.~\eqref{odd_T} to tensors on odd sites.

The decompositions are approximated by restricting the range of the summation for the new index $m$
as $1\le m \le D$~\footnote{
  Of course one can arbitrarily choose the truncation order of the new index at each coarse-grained step.
  In this paper, however, we keep it to $D$ for simplicity.
  This choice is convenient in actual computations because
  the bond dimension of the tensor does not change in every coarse-graining step.
}
\begin{align}
  \label{even_T_D}
  T_{ijkl} \approx \sum_{m=1}^{D} S^{[1]}_{(ij)m} \sigma^{[13]}_{m} S^{[3]}_{m(kl)},
  \\
  \label{odd_T_D}
  T_{ijkl} \approx \sum_{m=1}^{D} S^{[2]}_{(li)m} \sigma^{[24]}_{m} S^{[4]}_{m(jk)},
\end{align}
and the following coarse-grained tensor is obtained:
\begin{align}
  \label{eq:newtensor}
  T^{\mathrm{new}}_{ijkl}
  = \sqrt{\sigma^{[13]}_{i} \sigma^{[24]}_{j} \sigma^{[13]}_{k} \sigma^{[24]}_{l}}
  \sum_{a, b, c, d = 1}^{N} S^{[3]}_{i(cd)} S^{[4]}_{j(da)} S^{[1]}_{(cd)k} S^{[2]}_{(bc)l}.
\end{align}
In the first few steps in the TRG, $N^2$ might not be larger than $D$. In this case
$D$ is replaced by $N^{2}$ for $N^{2} < D$.~\footnote{
  In the case of $N^{2}<D$,  eqs.~(\ref{even_T_D})  and  (\ref{odd_T_D})  are not approximations but the same as
  eqs.~(\ref{even_T})  and  (\ref{odd_T}), respectively.
}
Note that the coarse-grained tensor $T^{\mathrm{new}}$ is defined by gathering $S^{[i]}$ ($i=1, 2, 3, 4$)
and its bond dimension is at most $D$.
One can see that the tensor network of $T$ is approximately expressed as a coarse network of $T^{\mathrm{new}}$.

The number of tensors decreases at each coarse-graining step since $T$ provides two $S$'s while $T^{\mathrm{new}}$ is made of four $S$'s.
Repeating the procedure above over and over again, we finally find that the network is expressed as a single tensor whose
indices are contracted with itself. The value of the tensor network is thus obtained.

Now let us turn to the coarse-graining of the tensor network with an impurity tensor.
As explained below, an important point is
that the number of impurity tensors does not increase beyond four (the impurity tensors are
located in four corners of a plaquette)
even if the coarse-graining step is repeated many times.

\begin{figure}[htbp]
  \centering
  \includegraphics[width=\hsize]{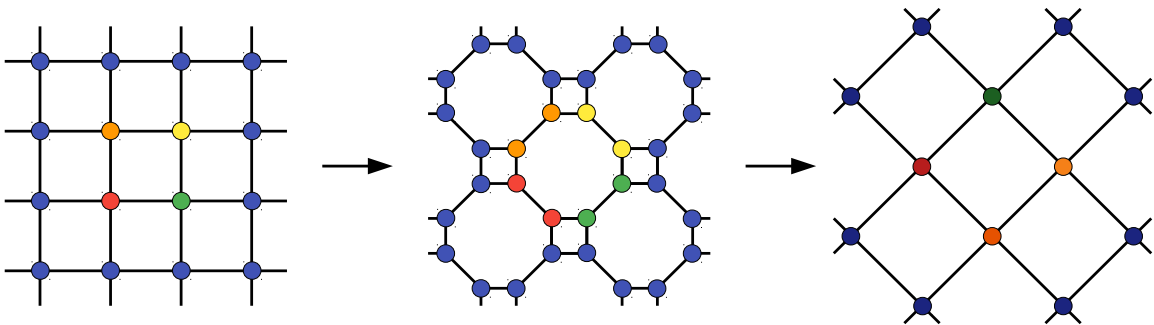}
  \caption{
    Coarse-graining of tensor network with four impurity tensors.
    Red, orange, yellow and, green circles denote the impurity tensors, and the normal tensors are shown by blue ones.
    Note that the impurity tensors colored in red, orange, yellow, and green are inherited in both decomposition and contraction processes.
    A key point is that the number of impurity tensors does not change through the coarse-graining.
  }
  \label{fig:TRG_Z1}
\end{figure}

Suppose we consider a coarse-graining procedure for the tensor network with four impurity tensors
as shown in figure~\ref{fig:TRG_Z1} (left),
where the impurity tensors are represented as the red, orange, yellow, and green circles while
the normal tensors are given by the blue circles.
In the first step, the rank-four tensors are decomposed into rank-three tensors.
As seen in the middle panel in figure~\ref{fig:TRG_Z1}, those rank-three impurity tensors form a closed loop.
Thanks to this structure, after taking the contraction, one obtains a network
which again contains only four impurity tensors as shown in the right panel in figure~\ref{fig:TRG_Z1}.
In this way, one can suppress the spread of the impurity tensors.

Starting from a tensor network with a single impurity tensor such as eq.~\eqref{eq:Z1_tensornetwork},
one obtains a network with two neighboring impurity tensors after a course-graining step. In the next step it
is coarse-grained to yield a network with three impurity tensors located on three corners of a plaquette.
In this way, the number of impurity tensors stepwisely increases until it reaches four.
After that, as shown above, the number of impurity tensors does not change anymore.

\section{Systematic error from discretization of scalar fields}
\label{sec:Kdependence}

As shown in section~\ref{sec:TN_2Dphi4} we discretize the scalar fields using the Gauss--Hermite quadrature
in order to create a finite dimensional tensor, which gives the approximated expressions for
the partition function and expectation values.
The order of approximation is characterized by the degree of the Hermite polynomial $K$.
In this appendix we discuss the systematic error associated with the approximation, namely the finite $K$-effect,
for the expectation value of the field.~\footnote{In ref.~\cite{Kadoh:2018hqq} we also checked the $K$-dependence of the free energy in the free boson system (with the Wilson term),
  and the reader can refer to the paper for more discussion.}

For an one dimensional integral in eq.~\eqref{eq:GHquadrature},  the Gauss--Hermite quadrature gives an exact result for $2K - 1$ degree (or less) polynomial target functions.
In the case of multi dimensional integrals, unfortunately, the convergence is not trivial.
We have to check whether or not the numerical results do not have large systematic errors from the finite $K$-effect with $K=256$, which is employed in this study.

\begin{figure}[htbp]
  \centering
  \includegraphics[width=0.8\hsize]{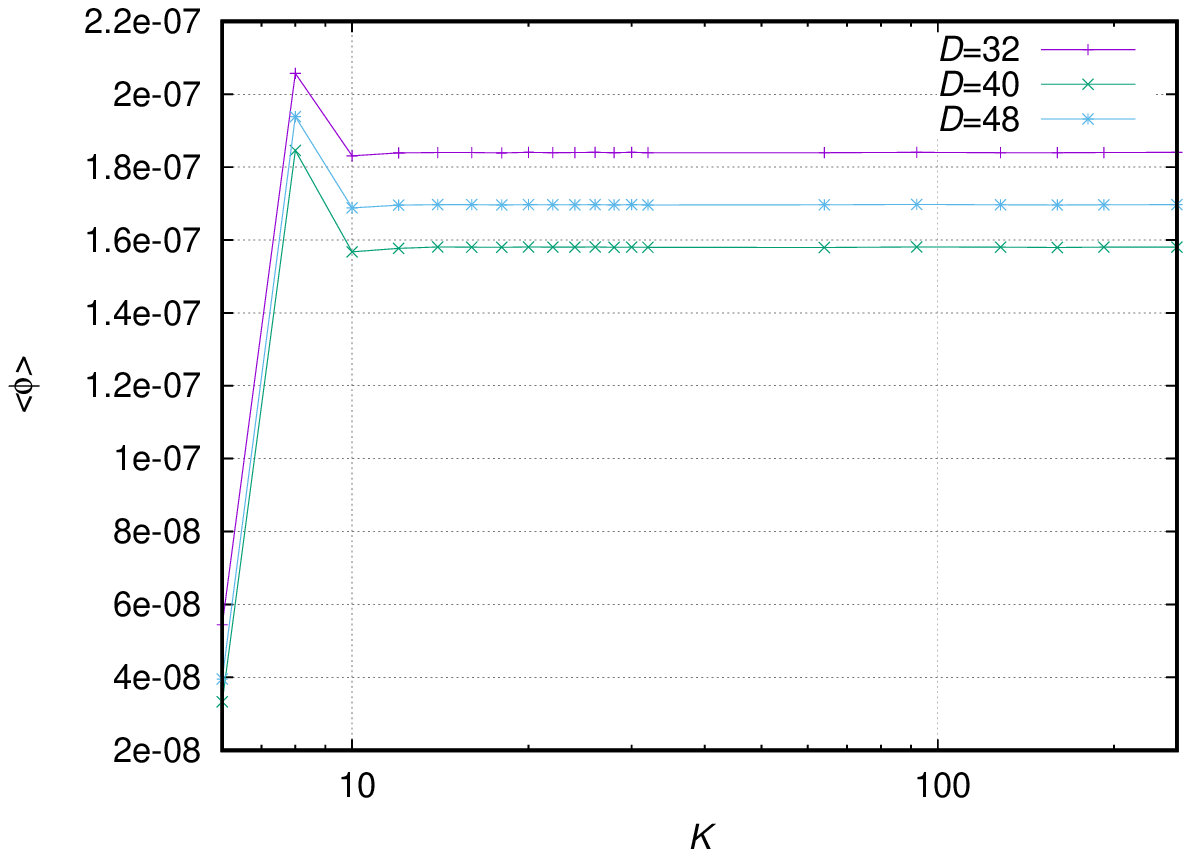}
  \caption{$K$-dependence of $\left< \phi \right>$ at $\mu_{0}^{2} =\mu_{0, \mathrm{rep.}}^{2}$ (see eq.~\eqref{eq:mu2_0rep}), $\lambda = 0.05$, $h=10^{-12}$, and $L=1024$.}
  \label{fig:phi_K2-256_D32-48}
\end{figure}

In figure~\ref{fig:phi_K2-256_D32-48} the $K$-dependence of  $\left< \phi \right>$ is presented.　It is clear that the results do not depend on $K$ for $K\in \left[ 64,256 \right]$ while much larger $D$-dependence
is observed.
Thus we can conclude that our choice $K=256$ is large enough, and the main source of systematic errors is the
truncation of the SVD for constructing the initial tensor and the coarse-graining steps in the TRG.

\acknowledgments

This work was supported by
the Ministry of Education, Culture, Sports, Science and Technology (MEXT) as ``Exploratory Challenge on Post-K computer'' (Frontiers of Basic Science: Challenging the Limits),
and JSPS KAKENHI Grant Numbers JP16K05328, JP17K05411, JP18J10663.

\providecommand{\href}[2]{#2}\begingroup\raggedright\endgroup

\end{document}